\documentclass[12pt,aps,showpacs]{revtex4}

\newcommand{\beeq}{\begin{equation}}
\newcommand{\eneq}{\end{equation}}
\newcommand{\be}{\begin{eqnarray}}
\newcommand{\ee}{\end{eqnarray}}
\newcommand{\bpic}{\begin{picture}}
\newcommand{\epic}{\end{picture}}
\newcommand{\bs}{\begin{scriptsize}}
\newcommand{\es}{\end{scriptsize}}

\def\la{\raise.16ex\hbox{$\langle$} \, }
\def\ra{\, \raise.16ex\hbox{$\rangle$} }

\def\d{\delta}
\def\e{\epsilon}

\def\G{\Gamma}
\def\r{\rho}

\def\D{\Delta}
\def\L{\Lambda}
\def\l{\lambda}

\def\Box{\kern1pt\vbox{\hrule height 1.2pt\hbox{\vrule width 1.2pt\hskip 3pt
   \vbox{\vskip 6pt}\hskip 3pt\vrule width 0.6pt}\hrule height 0.6pt}\kern1pt}
\def\gtwid{\mathrel{\raise.3ex\hbox{$>$\kern-.75em\lower1ex\hbox{$\sim$}}}}
\def\ltwid{\mathrel{\raise.3ex\hbox{$<$\kern-.75em\lower1ex\hbox{$\sim$}}}}

\begin{document}

\vskip -0.2in

\rightline{LPT-ORSAY 04-47, UFIFT-QG-04-1}

\vskip 0.5in

\title{Quantum effects can render $w<-1$ on cosmological scales}

\author{V. K. Onemli$^{\dagger}$} \author{R. P.
Woodard$^\ddagger$} \affiliation{${}^\dagger$LPT, Universit\'e
Paris Sud, B\^atiment 210, 91405 Orsay
France}\affiliation{${}^\ddagger$Department of Physics, University
of Florida, Gainesville, Florida 32611, USA}

\begin{abstract}
We report on a revision of our previous computation of the
renormalized expectation value of the stress-energy tensor of a
massless, minimally coupled scalar with a quartic self-interaction
on a locally de Sitter background. This model is important because
it demonstrates that quantum effects can lead to violations of the
weak energy condition on cosmological scales --- on average, not
just in fluctuations --- although the effect in this particular
model is far too small to be observed. The revision consists of
modifying the propagator so that dimensional regularization can be
used when the dimension of the renormalized theory is not four.
Although the finite part of the stress-energy tensor does not
change (in $D=4$) from our previous result, the counterterms do.
We also speculate that a certain, finite and separately conserved
part of the stress tensor can be subsumed into a natural
correction of the initial state from free Bunch-Davies vacuum.
\end{abstract}

\pacs{98.80.Cq, 04.62.+v}

\maketitle \vskip 0.2in

Caldwell \cite{Caldwell} was the first to point out that the
original supernova acceleration data \cite{SN1,SN2} are consistent
with a dark energy equation of state $w \equiv p/\rho$ less than
minus one, which would violate the weak energy condition.
Subsequent analyses of better and more abundant data have
confirmed this possibility in the context of an evolving dark
energy equation of state whose current value is less than minus
one \cite{ASSS,CP,WM,NP,G,FWZ,CKPCB}. However, it should be noted
that realizing this possibility generally implies accepting a
somewhat low value for the current Hubble parameter and a somewhat
high value for fraction of the critical density currently
comprised by dark matter \cite{ASS}. When combined data sets are
used, which restrict these two parameters, the data are well fit
by a simple cosmological constant with $w = -1$ \cite{WT,ASS,JBP}.

If the current phase of acceleration is actually driven by dark
energy which violates the weak energy condition it would pose an
excruciating problem for fundamental theory because the universe
has existed over 13 Gyr
\cite{CHT,CJM,HJW,F,SCK,PZ,ST,J,PZ2,LL,NO1,BW,HL1,SSD,NO2,NO3,HL2,HL3,MPW,M,MW,NOO,VS}.
One can get $w<-1$ by using scalars with a negative kinetic term,
however, such models are unstable against the production of
positive-negative energy particles. This instability obviously
grows worse as the negative energy particle is endowed with
interactions with more species of positive energy particles. The
minimal case is for it to interact only with gravity. For a
specific model of this type Carroll, Hoffman and Trodden
\cite{CHT} estimated that such a scalar would decay into two
gravitons and three scalars over the lifetime of the universe
unless the interaction is cut off, by fiat, at about $100~{\rm
MeV}$. A more stringent and model-independent bound was obtained
by Cline, Jeon and Moore \cite{CJM} by considering the process
whereby a graviton loop produces two scalars and two photons in
empty space. They conclude that the diffuse gamma ray background
will be too high unless the interaction is cut off at about
$3~{\rm MeV}$. More recently Hsu, Jenkins and Wise have shown
\cite{HJW} that instabilities occur in any scalar theory which
exhibits $w<-1$, irrespective of how this is achieved. Clearly,
the observed persistence of the universe can only be consistent
with a relatively brief phase of $w<-1$.

One way to achieve such a self-limiting phase --- without violating
classical stability --- is through quantum effects. Four years before
the first supernova data appeared
Starobinsky and Yokoyama studied a model which does this \cite{SY}. It
consists of a massless, minimally coupled scalar with a quartic
self-interaction which is released in free Bunch-Davies vacuum on a
locally de Sitter background. By applying Starobinsky's technique of
stochastic inflation \cite{Staro}, they were able to show that the
scalar initially moves up its potential, which would violate the weak
energy condition by increasing the Hubble parameter. Eventually the
upward push from inflationary particle production is compensated by
the downward classical force and the Hubble parameter asymptotes
to a constant value. The time for the process goes like the inverse
square root of the coupling constant.

The solution of Starobinsky and Yokoyama \cite{SY} is
nonperturbative, but it includes only the leading logarithms of
the scale factor at each order. (We thank A. A. Starobinsky for
pointing this out.) One can see that the vacuum energy increases
this way, but it is not possible to either verify stress-energy
conservation or to directly check that $\rho + p$ is negative. We
recently computed the fully renormalized expectation value of this
model's stress-energy tensor at one and two loops \cite{OW}.
Although our analysis was explicitly perturbative it produced the
complete result at one and two loop orders, thereby allowing
verification of conservation and a direct check that $\rho + p$ is
in fact negative.

What made our calculation possible was a relatively simple form for the
$D$-dimensional scalar propagator, which allowed us to employ dimensional
regularization. The scalar propagator is constrained to obey the equation,
\begin{equation}
\frac{\partial}{\partial x^{\mu}} \Bigl( \sqrt{-g(x)} g^{\mu\nu}(x)
\frac{\partial}{\partial x^{\nu}} i \Delta(x;x') \Bigr) = i \delta^D(x-x')
\; . \label{propeqn}
\end{equation}
Were de Sitter invariance maintained one could express $i\Delta(x;x')$
entirely in terms of the geodesic length $\ell(x;x')$. However, Allen
and Follaci long ago showed that the massless, minimally coupled scalar
possesses no normalizable, de Sitter invariant states \cite{AF}. We
chose to introduce the inevitable breaking of de Sitter invariance in
a manner consistent with the homogeneity and isotropy of cosmology.
In our conformal coordinate system the invariant element is,
\begin{equation}
g_{\mu\nu} dx^{\mu} dx^{\nu} = a^2(\eta) \Bigl(-d\eta^2 + d\vec{x}
\! \cdot \! d\vec{x} \Bigr) \;\;\; , \qquad a(\eta) \equiv
-\frac1{H \eta} \qquad ,
\end{equation}
where $\Lambda = (D-1) H^2$ relates the Hubble constant $H$ to the
cosmological constant $\Lambda$ in $D$ dimensions. Our solution to
(\ref{propeqn}) depends upon $a \equiv a(\eta)$ and $a' \equiv a(\eta')$
in addition to the de Sitter invariant length function,
\begin{equation}
y(x;x') \equiv 4 \sin^2\Bigl(\frac12 H \ell(x;x')\Bigr) = a a' H^2
\Bigl(\Vert \vec{x} - \vec{x}' \Vert^2 - (|\eta - \eta'|-i\d)^2
\Bigr) \; ,
\end{equation}
where $\d$ is a positive real number. We normalize the scale
factor to $a=1$ when the state is released, so that $a > 1$
throughout the computation.

Our previous computation \cite{OW} was based upon the solution,
\be
i\D_{\mbox
{\tiny{old}}}=\frac{H^{D-2}}{(4\pi)^\frac{D}{2}}\Bigg\{-\sum_{n=0}^{\infty}
\frac{1}{n-\frac{D}{2}+1}\frac{\G(n+\frac{D}{2})}{\G(n+1)}
\left(\frac{y}{4}\right)^{n-\frac{D}{2}+1}
+\frac{2^{4-D}}{2-\frac{D}{2}}\G(\frac{D}{2}+1)\nonumber\\+\frac{\G(\frac{D}{2})}{2^{D-4}}\sum_{n=1}^{\infty}
\frac{1}{n}\frac{\G(n+D-1)}{\G(n+\frac{D}{2})}
\left(\frac{y}{4}\right)^n+\frac{\G(D-1)}{2^{D-4}}\ln{(aa')}\Bigg\}\;
.\label{old}
\ee
It consists of four terms: (i) An infinite series of
$D$-dependent powers of $\frac{y}{4}$; (ii) A $D$-dependent
constant; (iii) An infinite series of integer powers of
$\frac{y}{4}$ and (iv) The $\ln{(aa')}$ term. The normalization of
(i) is set by getting the delta function. The constant term
(ii) is a homogeneous solution and can be added for free. It was
chosen to cancel the singularity at $D=4$ in the $n=1$ term of the
series (i). Although the sum of (i) and (ii) is finite for small
$y$, it diverges at $y=4$ and beyond. The point of the second
infinite series (iii) is to cancel this divergence at $D=4$. However,
the series (iii) does not solve the homogeneous equation. The de Sitter
breaking term (iv) must be added for this purpose.

In addition to the $\phi^4$ stress-energy tensor \cite{OW}, the old
propagator (\ref{old}) was used in \cite{POW1,POW2} to compute the
one loop vacuum polarization from scalar QED. While computing the one loop
self-energy of a Yukawa-coupled fermion \cite{PW1} the propagator
was recently modified to make it valid for regulating a theory
whose dimension will not ultimately be taken to $D=4$,
\be
i\D_{\mbox
{\tiny{new}}}=\frac{H^{D-2}}{(4\pi)^\frac{D}{2}}\Bigg\{-\sum_{n=0}^{\infty}
\frac{1}{n-\frac{D}{2}+1}\frac{\G(n+\frac{D}{2})}{\G(n+1)}
\left(\frac{y}{4}\right)^{n-\frac{D}{2}+1}
-\frac{\G(D-1)}{\G(\frac{D}{2})}\pi\cot{(\pi\frac{D}{2})}\nonumber\\
+\sum_{n=1}^{\infty}
\frac{1}{n}\frac{\G(n+D-1)}{\G(n+\frac{D}{2})}
\left(\frac{y}{4}\right)^n+\frac{\G(D-1)}{\G(\frac{D}{2})}\ln{(aa')}\Bigg\}\; .
\label{new}
\ee
This change makes a few insignificant alterations in the finite part of
the one loop vacuum polarization \cite{PW2}. The purpose of this brief
report is to consider possible changes from using the modified propagator
to re-compute the expectation value of the $\phi^4$ stress-energy tensor.
We also take this opportunity to correct a minor error --- in the
normalization of the de Sitter breaking term (iv) --- in the previously
published expressions for the new propagator \cite{PW1,PW2}. This error
has no effect on the one loop computations for which the new propagator
was previously employed but it becomes quite significant at higher loops.

When the new propagator (\ref{new}) is employed (with $D=4-\e$) to
re-compute the expectation value of the $\phi^4$ stress-energy tensor
we find the same fully renormalized result but totally
different results for the mass-squared, the conformal and the
cosmological constant counterterms. We report the various changes below,
giving the previously reported results \cite{OW} with subscript
``old'', followed by the new results with subscript ``new'',
\be
\d m^2_{\mbox
{\tiny{old}}}&=&-\frac{\l H^{2-\e}}{2^4\pi^{2-{\frac{\e}{2}}}}
\frac1{\e} \G(3 - \frac{\e}2) +O(\l^2)
\nonumber\\&\longrightarrow&  -\frac{\l
H^{2-\e}}{2^{5-\e}\pi^{2-{\frac{\e}{2}}}} \frac{\G(3
-\e)}{\G(2-\frac{\e}{2})}\pi\cot(\frac{\pi}{2}\e) +O(\l^2)\equiv\d
m^2_{\mbox {\tiny{new}}} ,\ee \be \d\xi_{\mbox {\tiny{old}}} &=&
\frac{\l}{2^4\pi^2} \left(\frac{H^2}{\pi} \right)^{\frac{\e}{2}}
\Bigg\{ \frac{\zeta_{\mbox {\tiny{old}}}}{2\e\G(4 - \e)} +
\d\xi^{\mbox
{\tiny{old}}}_{\mbox{\tiny{fnt}}}\Bigg\}\nonumber\\&\longrightarrow&
\frac{\l}{2^4\pi^2} \left(\frac{H^2}{4\pi} \right)^{\frac{\e}{2}}
\Bigg\{ \frac{\G(2-\frac{\e}{2})}{\G(4-\e)}\frac{\zeta_{\mbox
{\tiny{new}}}}{2\e} + \d\xi^{\mbox
{\tiny{new}}}_{\mbox{\tiny{fnt}}}\Bigg\}\equiv{\delta\xi_{\mbox
{\tiny{new}}}} \; , \ee where \be \zeta_{\mbox
{\tiny{old}}}&=&\left(\frac{\pi}{\mu H} \right)^\e \left(1 -
\frac{\e}2\right)^2 \G(1 - \e) \G\left(1 -
\frac{\e}2\right)\nonumber\\&\longrightarrow&
\left(\frac{2\pi}{\mu H} \right)^\e \left(1 -
\frac{\e}2\right)\G(1 - \e)\equiv\zeta_{\mbox {\tiny{new}}}\; ,\ee
and\be \frac{\d\Lambda_{\mbox {\tiny{old}}}}{8\pi G} &=& \frac{\l
H^4}{2^6\pi^4} \Bigg\{\left( \frac{\pi}{ H^2} \right)^{\e}
\frac{\G^2(3-\frac{\e}{2})}{8 \e^2} - \frac{\zeta_{\mbox
{\tiny{old}}}}{2\e^2} \frac{\G(3-\frac{\e}{2})}{\G(3-\e)}
-\frac{\d\xi^{\mbox {\tiny{old}}}_{ \mbox{\tiny{fnt}}}}{\e}\left(3
- \e\right) \G\left(3 - \frac{\e}2\right) +
\d\Lambda_{\mbox{\tiny{fnt}}}^{\mbox
{\tiny{old}}}\Bigg\}\nonumber\\
&\longrightarrow& \frac{\l H^4}{2^6\pi^4} \Bigg\{\left(
\frac{4\pi}{H^2} \right)^{\e}
\frac{\G^2(3-\e)}{\G^2(2-\frac{\e}{2})}\frac{\pi^2}{32}\cot^2{(\frac{\pi}{2}\e)}-\frac{\zeta_{\mbox
{\tiny{new}}}}{4\e}\pi\cot{(\frac{\pi}{2}\e)}\nonumber\\
&&-\d\xi^{\mbox
{\tiny{new}}}_{\mbox{\tiny{fnt}}}\frac{\G(4-\e)}{\G(2-\frac{\e}{2})}\frac{\pi}{2}\cot{(\frac{\pi}{2}\e)}
+ \d\Lambda^{\mbox
{\tiny{new}}}_{\mbox{\tiny{fnt}}}\Bigg\}\equiv\frac{\d\Lambda_{\mbox
{\tiny{new}}}}{8\pi G} \; .
\ee
We make the same choices as before for the arbitrary finite parts
of the cosmological and conformal counterterms,
\be
\d\xi^{\mbox{\tiny{new}}}_{\mbox{\tiny{fnt}}}&\equiv&
-\frac{7}{36} + \frac1{12} \ln\left(\frac{2\mu}{H}
\right)=\d\xi^{\mbox{\tiny{old}}}_{\mbox{\tiny{fnt}}}\nonumber\\
\d\Lambda^{\mbox{\tiny{new}}}_{\mbox{\tiny{fnt}}}&\equiv&
\frac1{18} - \frac{\pi^2}{12}=
\d\L^{\mbox{\tiny{old}}}_{\mbox{\tiny{fnt}}} \; .
\ee
The renormalized energy density and pressure are unchanged from their
previous values,
\begin{eqnarray}
\r_{\mbox{\tiny{ren}}} & = & \frac{\Lambda}{8\pi G} + \frac{\l H^4}{2^6\pi^4}
\Bigg\{ \frac12 \ln^2\left(a\right) + \frac29 a^{-3} - \frac12
\sum_{n=1}^{\infty}
\frac{n+2}{(n+1)^2} a^{-n-1} \Bigg\} + O(\lambda^2) \; , \label{rho} \\
p_{\mbox{\tiny{ren}}} & = & - \frac{\Lambda}{8\pi G} - \frac{\l H^4}{2^6\pi^4}
\Bigg\{\frac12 \ln^2\left(a\right) + \frac13 \ln\left(a\right) +
\frac16 \sum_{n=1}^{\infty} \frac{n^2-4}{(n+1)^2}
a^{-n-1}\Bigg\} + O(\lambda^2) \; . \label{pres}
\end{eqnarray}
Hence
\be
\rho_{\mbox{\tiny{ren}}} + p_{\mbox{\tiny{ren}}} =
\frac{\l H^4}{2^6\pi^4} \Bigg\{- \frac13 \ln\left(a\right) +
\frac29 a^{-3} - \frac16 \sum_{n=1}^{\infty} \frac{n+2}{n+1}
a^{-n-1}\Bigg\} + O(\lambda^2) \; , \label{it}
\ee
violates the weak energy condition on cosmological scales.

Before concluding we wish to make three comments. First, $w + 1$
is unobservably small in this model. From (\ref{rho}-\ref{pres})
we compute, \be w \equiv
\frac{p_{\mbox{\tiny{ren}}}}{\r_{\mbox{\tiny{ren}}}} =
-\left\{1+\frac{\l H^2 G}{(2\pi)^3}\left[\frac{1}{9}\ln{(a)} +
O(a^{-2})\right]+O(\l^2)\right\} \; . \ee Using $H_0 \simeq
71~{\rm km/(s\cdot Mpc)}$ one finds the dimensionless number $G
H_0^2 \equiv G H_0^2 (\hbar/c^5) \simeq 1.5 \times 10^{-122}$. One
might hope this minuscule prefactor could be enhanced by the
coupling constant $\lambda$ or by the secular factor of $\ln(a)$.
However, our analysis has been perturbative --- which rules out
$\lambda > 1$ --- and the data shows that acceleration only began
at about $z \simeq 1$ --- which means that $a_0 \simeq 2$ if we
assume the process began when the deceleration became negative. In
any case, the nonperturbative solution of Starobinsky and Yokoyama
\cite{SY} shows that $w$ approaches $-1$ after $\ln(a) \simeq
1/\sqrt{\lambda}$, so the weak energy condition is never violated
by very much in this model. What the model does establish, in a
simple setting and beyond the point of dispute, is that quantum
effects can induce a self-limiting phase in which a classically
stable theory violates the weak energy condition on cosmological
scales. Once this is accepted one can search for other models in
which the effect may be observable. Such a model has been proposed
by Parker and Raval \cite{PR1,PR2}, and slightly modified by
Parker and Vanzella \cite{PV}.

Our second comment concerns the exponentially falling portions of the
stress-energy tensor,
\begin{eqnarray}
\r_{\mbox{\tiny{falling}}} & \equiv & \frac{\lambda H^4}{2^7 \pi^4}
\Biggl\{\frac49 a^{-3} - \sum_{n=1}^{\infty} \frac{n+2}{(n+1)^2} a^{-n-1}
\Bigg\} \; , \\
p_{\mbox{\tiny{falling}}} & \equiv & - \frac{\lambda H^4}{2^7 \, 3
\pi^4} \sum_{n=1}^{\infty} \frac{n^2 - 4}{(n+1)^2} a^{-n-1} \; .
\end{eqnarray}
Note that these terms are separately conserved,
\begin{equation}
\dot{\r}_{\mbox{\tiny{falling}}} = - 3 H ( \r_{\mbox{\tiny{falling}}} +
p_{\mbox{\tiny{falling}}} ) \; .
\end{equation}
We conjecture that these terms can be subsumed into a modification
of the initial free Bunch-Davies vacuum at $a=1$. Even in flat
space one can see that the free state wave functional,
\begin{equation}
\Omega\Bigl[\phi\Bigr] = N \exp\left[-\frac12 \int d^3x \, \phi(\vec{x})
\sqrt{-\nabla^2} \phi(\vec{x}) \right] \; ,
\end{equation}
must suffer nonlocal corrections of order $\lambda \phi^4$. We propose
that using this perturbatively corrected initial state would cancel
the falling portions of the stress-energy leaving only the infrared
logarithms,
\begin{eqnarray}
\r_{\mbox{\tiny{conj}}} & = & \frac{\Lambda}{8\pi G} + \frac{\l H^4}{2^6\pi^4}
\Bigg\{ \frac12 \ln^2\left(a\right) \Bigg\} + O(\lambda^2) \; , \\
p_{\mbox{\tiny{conj}}} & = & - \frac{\Lambda}{8\pi G} - \frac{\l H^4}{2^6\pi^4}
\Bigg\{\frac12 \ln^2\left(a\right) + \frac13 \ln\left(a\right)\Bigg\} +
O(\lambda^2) \; .
\end{eqnarray}

Our final comment is that quantum fluctuations of the
stress-energy operator will of course violate the weak energy
condition for a classical background such as de Sitter which is
right on the boundary $\rho + p = 0$ \cite{Win,Vac}. The model we
have considered gives a more serious violation, in the {\it
average value} of the stress-energy tensor, rather than in
fluctuations about an average which obeys the condition.

\end{document}